\documentclass[aps,twocolumn,showpacs,amsmath,amssymb]{revtex4}
\usepackage{graphicx,color}
\usepackage{bm}
\bmdefine{\bk}{k}
\bmdefine{\bx}{x}
\bmdefine{\br}{r}

\newcommand{\rb}{\bar{\rho}}
\newcommand{\lka}{\left(}
\newcommand{\rka}{\right)}

\newcommand{\ldka}{\left[}
\newcommand{\rdka}{\right]}
\newcommand{\dbk}{\delta_\bk}
\begin{document}

\title{Galaxy clustering constraints on deviations from Newtonian 
gravity at cosmological scales}

\author{Akihito Shirata$^{(1)}$, Tetsuya Shiromizu$^{(1,3,4)}$ Naoki
Yoshida$^{(2)}$, and Yasushi Suto$^{(3)}$}

\affiliation{$^{(1)}$Department of Physics, Tokyo Institute of Technology, 
Tokyo 152-8551, Japan}

\affiliation{$^{(2)}$ Department of Physics and Astrophysics, 
Nagoya University, Nagoya 464-8602, Japan}

\affiliation{$^{(3)}$Department of Physics, The University of Tokyo,  Tokyo
113-0033, Japan}

\affiliation{$^{(4)}$Advanced Research Institute for Science and Engineering, 
Waseda University, Tokyo 169-8555, Japan}

\date{\today}

\begin{abstract}
In spite of the growing observational evidence for dark matter and dark
energy in the universe, their physical nature is largely unknown. In
fact, several authors have proposed modifications of Newton's law of
gravity at cosmological scales to account for the apparent acceleration
of the cosmic expansion. Inspired by such suggestions, we attempt to
constrain possible deviations from Newtonian gravity by means of the
clustering of SDSS (Sloan Digital Sky Survey) galaxies. To be specific,
we assume a simple modification of Newton's law with an additional
Yukawa-type term characterized by the amplitude $\alpha$ and the length
scale $\lambda$.  Adopting spatially-flat universes dominated by cold
dark matter and/or dark energy, we solve a linear perturbation equation
for the growth of density fluctuations. In particular, we find an exact
analytic solution for the Einstein -- de Sitter case.  Following the
Peacock-Dodds prescription, we compute the nonlinear power spectra of
mass fluctuations, perform a statistical comparison with the SDSS galaxy
data, and derive constraints in the $\alpha$-$\lambda$ plane; for
instance, we obtain the constraints of $-0.5<\alpha <0.6$ and
$-0.8<\alpha<0.9$ (99.7\% confidence level) for $\lambda=5h^{-1}$Mpc and
$10h^{-1}$Mpc, respectively.  We also discuss several future
possibilities for improving our analysis.
\end{abstract}


\maketitle


\label{sec:introduction}
\section{Introduction}

 Recent observations on cosmological scales suggest the presence of a
non-zero cosmological constant or dark energy in the universe.  Adopting
spatially-flat Friedmann-Robertson-Walker (FRW) models, type Ia
supernovae \cite{Knop:2003iy} and cosmic microwave background (CMB) data
\cite{Bennett:2003bz,Spergel:2003cb} are simultaneously well fit by
$\Omega_\Lambda \approx 0.7$ and $\Omega_{\rm m} \approx 0.3$, where
$\Omega_\Lambda$ and $\Omega_{\rm m}$ denote the dimensionless
cosmological constant and the density parameter of non-relativistic
matter, respectively.  In fact, nearly all the available observational
data are in good agreement with the above model parameters.

 On the other hand, the physical origin and nature of such dark energy
components remain to be understood.  In this paper we will consider a
possible deviation from Newton's law of gravity that are motivated by
recent proposals for an alternative to dark energy
\cite{Dvali:2000hr,Deffayet:2000uy} or as a possible consequence of dark
energy \cite{Sealfon,Nusser04,Farrar-Peebles}.  While this idea seems
exotic, the validity of Newton's law has not been demonstrated
rigorously on cosmological scales \cite{Adelberger:2003zx,Hoyle:2004cw}.
In addition, the presence of dark energy in the standard FRW model is
equivalent to introducing a repulsive force in the cosmic expansion.  In
a sense, this is equivalent to changing Newton's law of gravity on
cosmological scales by modifying the matter content of the universe.
The idea behind this paper is to explore observable consequences of a
modification of gravity itself.  We note that in retrospect this is
closer to Einstein's original idea in introducing the cosmological
constant; he introduced the $\Lambda$-term on the left-hand side of the
field equation.  Nowadays this term is usually interpreted as an
additional term in the energy-momentum tensor on the right-hand side.

 While our current approach is entirely empirical, there have been
several specific attempts to construct self-consistent cosmological
models including deviations from Newton's law on cosmological scales.
For instance, Dvali, Gabadadze and Porrati proposed a scenario (DGP
model, hereafter) \cite{Dvali:2000hr,Deffayet:2000uy} to explain the
accelerating universe as a result of gravity leaking into extra
dimensions in the context of a braneworld model.  According to the DGP
model, the accelerating universe is naturally explained without dark
energy component, but rather by the modification of Newton's law of
gravity on cosmological scales.  A variety of astrophysical and
cosmological consequences of the DGP model were discussed by several
authors \cite{Deffayet:2001pu,Lue:2004rj,
Alcaniz:2004kq,Lue:2004za}. Other proposals which suggest deviations
from Newton gravity on very large scales include a ``ghost condensation''
model \cite{Arkani-Hamed:2003uy,Arkani-Hamed:2003uz}.

We would like to derive constraints on deviations from Newton's law that
are independent of any specific models to the extent possible.  Thus, we
adopt an empirical parameterization of the deviation from Newton's law
over $10 \sim 100$ Mpc scales while the cosmic expansion is assumed to
exactly follow that of the standard FRW model.

Note that, strictly speaking, our approach is not fully self-consistent
in that our modified gravity law has an effective gravitational constant
$G_{\rm eff}=G_{\rm N}(1+\alpha)$ asymptotically in the large-scale
limit, but we still assume that the cosmic expansion follows the
standard FRW cosmology with $G_{\rm N}$.  As discussed below, however,
this does not significantly change our conclusions.

 Sealfon et al. \cite{Sealfon} recently carried out a similar analysis
in the same spirit.  They derive linear theory predictions for the
matter power spectrum in modified gravity models and obtain constraints
on deviations from the inverse-square law using 2dF (Two-degree Field)
and SDSS (Sloan Digital Sky Survey) galaxy data.  Our paper differs from
theirs in the following three respects:
 (a) We do not assume {\it a priori} that any fractional deviation from
 Newton's law is small.  The analysis of Sealfon et al. is applicable
 only in a perturbative regime (i.e., $\alpha \ll 1$).  (b) Sealfon et
 al. solved the linear perturbation equation for the growth of density
 fluctuations assuming an incorrect scaling between wave-number and the
 scale factor.  We, however, solve the original perturbation equation
 without such a scaling assumption.  Finally (c) we take into account
 gravitational non-linearity by applying the Peacock-Dodds
 prescription \cite{Peacock:1996ci}.
 As a result, we believe that the constraints in our paper are
more reliable.

 The rest of the paper is organized as follows.  Section II describes
our models for modified gravity.  In Sec. III we solve the linear
perturbation equation for density fluctuations in modified gravity
models, and the nonlinear power spectra are computed in Sec. IV by
applying the Peacock-Dodds prescription. Section V compares these
theoretical predictions with the observed power spectrum of SDSS
galaxies. Finally Sec. VI is devoted to conclusions and discussion.

\label{sec:Method}
\section{Model Assumptions}

 The basic equation that we solve in this paper is 
\begin{equation}
\label{eq:basiceq}
	\ddot{\delta} + 2 H(t) \dot{\delta} - \Delta_{\br} \Phi = 0 ,
\end{equation}
where 
\begin{equation}
\label{eq:delrho}
	\delta(\br,t) \equiv 
\frac{\rho(\br,t) - \bar{\rho}(t)}{\bar{\rho}(t)} 
\end{equation}
is a fractional mass density fluctuation, $\bar{\rho}(t)$ is the mean
mass density, $H(t)$ is the Hubble parameter, $\Phi$ is the modified
Newtonian potential and $\br$ is a proper coordinate.  The dot denotes
the derivative with respect to the cosmic time $t$.  To solve equation
\eqref{eq:basiceq} one has to specify the functional form of $H(t)$ and
$\Phi$.  In addition, we will confront theoretical predictions against
the observed power spectrum of SDSS galaxies.  Thus we need an
additional assumption of the spatial biasing of SDSS galaxies with
respect to the underlying dark matter.  We adopt the following four
major assumptions in performing the analysis.

(i) For the gravitational potential $\Phi$, we consider two models.  The
first (Model I) is to simply change the amplitude of Newton's constant,
$G_{\rm N}$:
\begin{equation}
\label{eq:ModelI}
\Phi (\br) = - G_{\rm N} (1 + \alpha_0) 
\int dr'^3 \frac{\rho(\br')}{|\br-\br'|} ,
\end{equation}
where $\alpha_0$ is a scale-independent parameter characterizing the
deviation.  The effective gravitational constant
in this model is simply $G_{\rm eff} = G_{\rm N} (1+ \alpha_0)$.
While this simple model is useful in understanding a basic outcome of
modified gravitational theories, it may be inconsistent with precise tests of
Newton's gravity on small scales.  Therefore we also consider another model
(Model II) where the deviation is restricted to scales larger than
$\lambda$:
\begin{equation}
\label{eq:ModelII}
\hspace*{-0.4cm}
\Phi (\br) = - G_{\rm N} \int dr'^3 \frac{\rho(\br')}{|\br-\br'|} 
\ldka 1 + \alpha \lka 1 - e^{- \frac{|\br - \br'|}{\lambda}} \rka
\rdka. 
\end{equation}
Note that $\lambda$ is defined in proper, rather than comoving, length,
and we consider $1 h^{-1} \text{Mpc} \le \lambda \le 30 h^{-1}
\text{Mpc}$ below.  Equation (\ref{eq:ModelII}) recovers the
conventional Newton potential for $r \ll \lambda$, and asymptotically
approaches Model I with $G_{\rm eff} = G_{\rm N} (1 + \alpha)$ for $r
\gg \lambda$.  The transition between these two regimes is described by
the Yukawa-like term but our result below is insensitive to that
particular choice.

(ii) We assume that general relativity is valid on the horizon scale,
i.e., that the cosmic expansion is described by the standard FRW model.
Strictly speaking, this assumption may not be fully consistent with our
modified gravity law on local scales (Models I and II).  However, our
primary interest in this paper is the matter clustering on scales of $1
h^{-1} \text{Mpc} \le \lambda \le 30 h^{-1} \text{Mpc}$, which should
not be affected by evolution of structure on much larger scales that are
in the linear regime.  In other words, even if we introduced another
cutoff in Model II so as to recover the Newton gravity at scales $\gg
100h^{-1} \text{Mpc}$, our results would hardly be affected.  This
explains why we use the FRW model in describing the cosmic expansion
while non-Newtonian gravity controls the dark matter clustering.  For
the same reason, we assume that our modification of gravity does not
change the angular power spectrum of the CMB.

In reality, we do not necessarily have to assume the validity of general
relativity on horizon scales;  the unperturbed model could differ from 
general relativity and still be consistent with observations. We will
defer exploring such possibilities for now and adopt this assumption
just for simplicity throughout the present analysis.

(iii) In addition, we assume that the universe is dominated by cold dark
matter (CDM) and/or the cosmological constant.  To be specific, we
consider two spatially flat models with and without the cosmological
constant: $\Omega_{\rm m} + \Omega_{\Lambda} = 1$ (Table I).  Hence we use the
conventional models for the Hubble parameter $H$ in equation
\eqref{eq:basiceq} and the matter transfer function \cite{Eisenstein:1997ik}.

\begin{table}
\caption{\label{tab:table1}Parameters of cosmological models}
\begin{ruledtabular}
\begin{tabular}{cccccc}
Model&$\Omega_{\rm m}$&$\Omega_\Lambda$&$\Omega_{\rm b}$
&$h$\footnotemark[1]
&$\sigma_8$\\
\hline
$\Lambda$-CDM&0.3&0.7&0.04&0.7&0.9\\
EdS-CDM&1.0&0.0&0.04&0.7&0.6\\
\end{tabular}
\footnotetext[1]
{$h$ is the Hubble constant in units of 100km s$^{-1}$ Mpc$^{-1}$}
\end{ruledtabular}
\end{table}

(iv) Finally, we assume that SDSS galaxies are fair tracers of the total
mass distribution, but allow for a scale-independent
linear bias, $b_*$, for $L_*$ galaxies.  We use the power spectrum of
SDSS galaxies, $P_{\rm gg}(k; L_*)$, which is corrected for 
luminosity-dependent bias and refers to the clustering of $L_*$ galaxies
\cite{Tegmark}. Then we attempt to constrain the parameter space of $\alpha$
and $\lambda$ ($\alpha_0$ in Model I) by fitting to
\begin{equation}
\label{eq:Linearbias}
P_{\rm gg}(k; L_*) = b_*^2 P_{\text{\rm mass}}(k; \alpha,\lambda). 
\end{equation}
The scale independence of the biasing parameter is a fairly conventional
assumption in the linear regime where the growth rate is scale
independent. Note, however, that our Model II evolves in a weakly
scale-dependent fashion even in linear theory (see Fig. \ref{fig:fig2}
below).  Nevertheless we assume that $b_*$ is independent of $k$ for
simplicity.


\begin{figure}
\begin{tabular}{c}
      \resizebox{80mm}{!}{\includegraphics{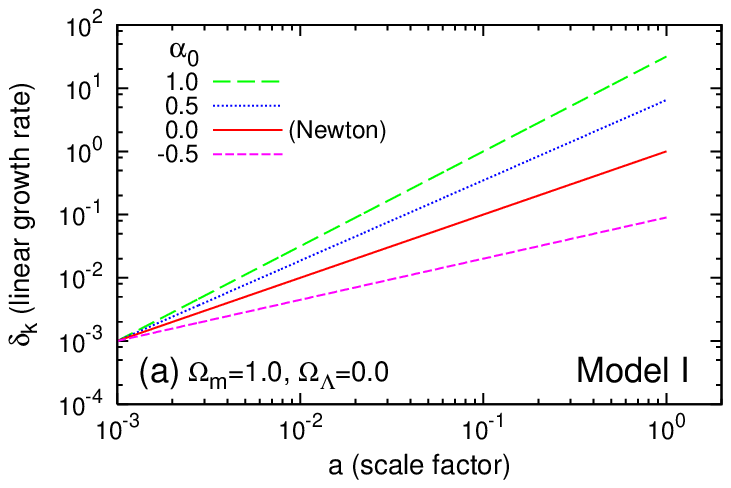}} \\
      \resizebox{80mm}{!}{\includegraphics{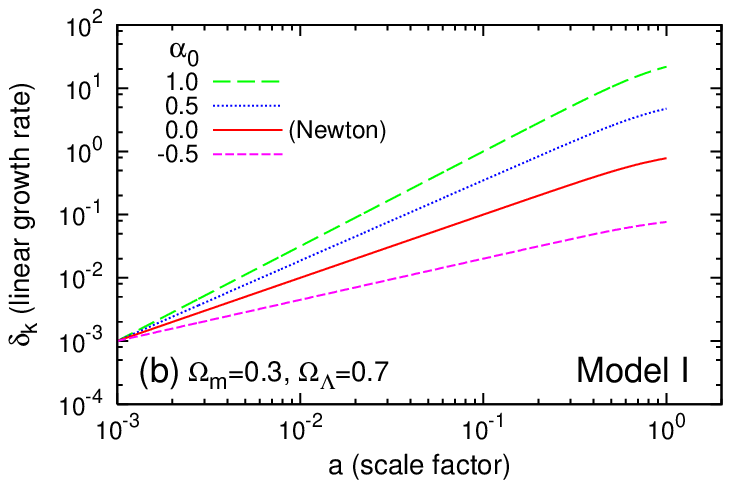}}
\end{tabular}
\caption{\label{fig:fig1} Linear growth rate in Model I where 
$G_{\rm eff}=(1
 + \alpha_0) G_{\rm N}$ is independent of scale; 
(a) $\Omega_{\rm m}=1$ and $\Omega_\Lambda=0$,
(b) $\Omega_{\rm m}=0.3$ and $\Omega_\Lambda=0.7$.
}
\end{figure}
\begin{figure*}
\begin{tabular}{cc}
      \resizebox{80mm}{!}{\includegraphics{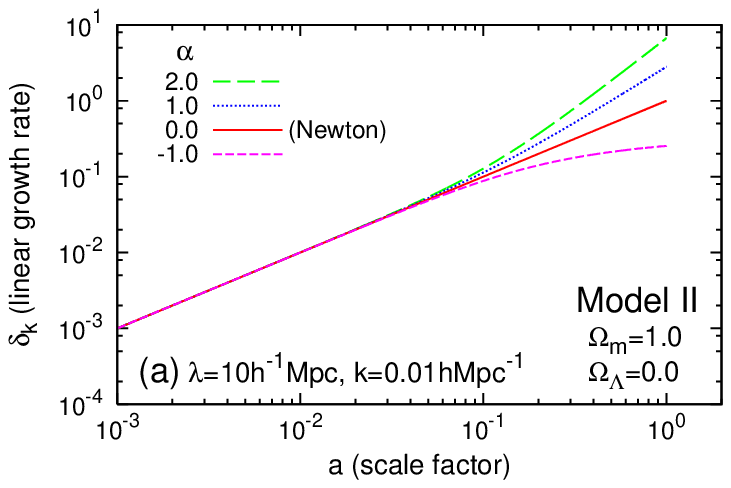}} &
      \resizebox{80mm}{!}{\includegraphics{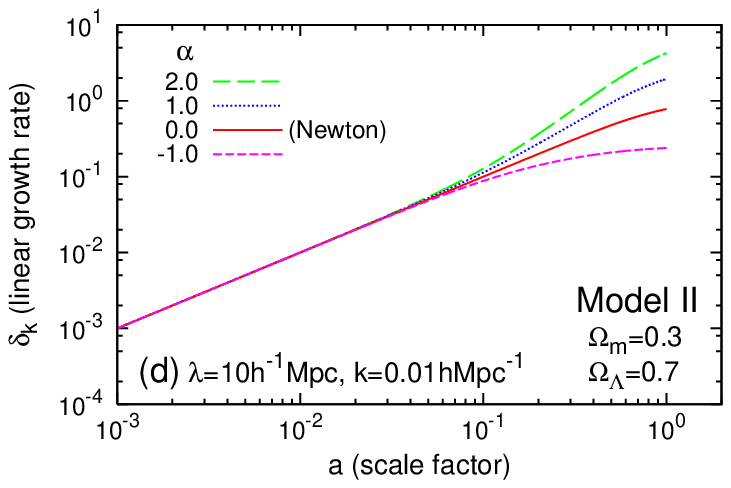}} \\
      \resizebox{80mm}{!}{\includegraphics{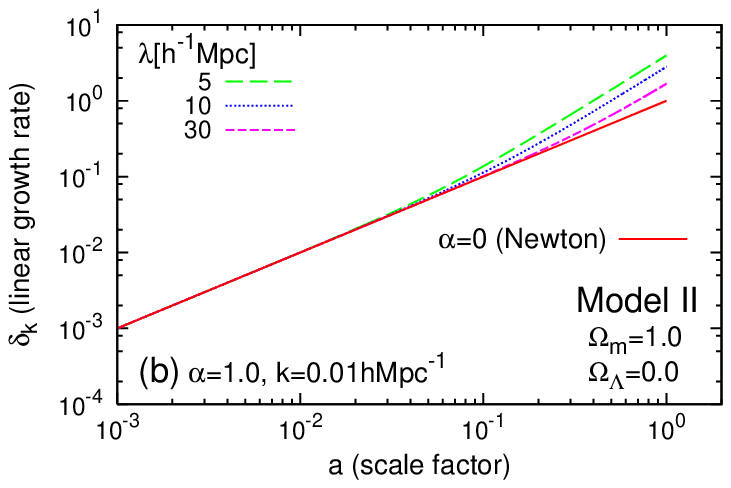}} &
      \resizebox{80mm}{!}{\includegraphics{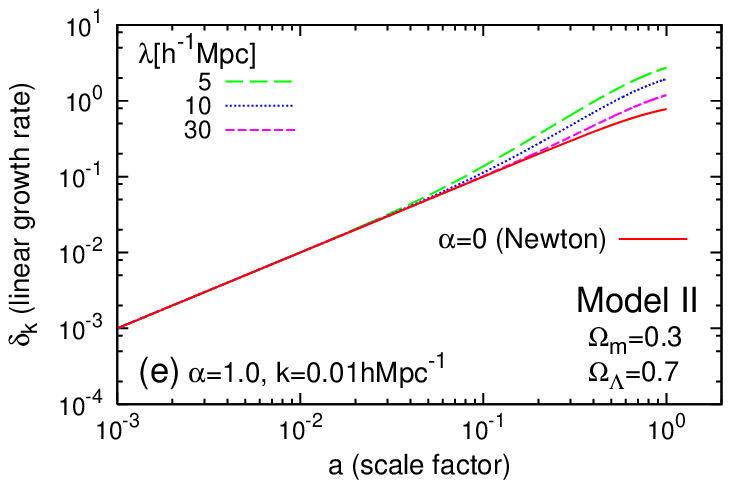}} \\
      \resizebox{80mm}{!}{\includegraphics{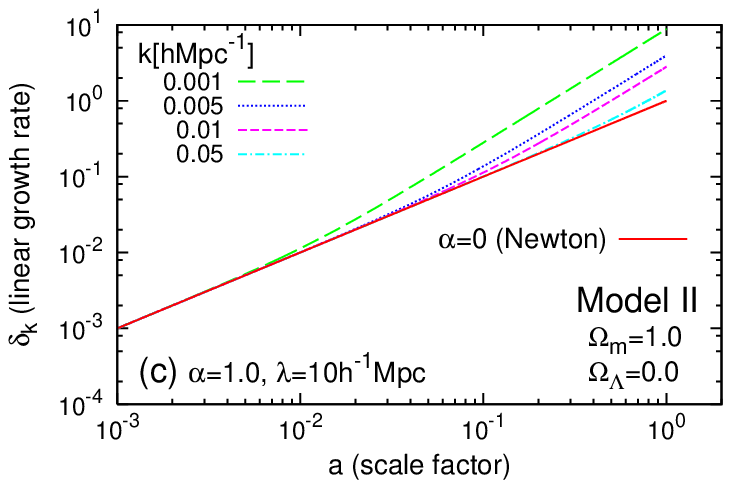}} &
      \resizebox{80mm}{!}{\includegraphics{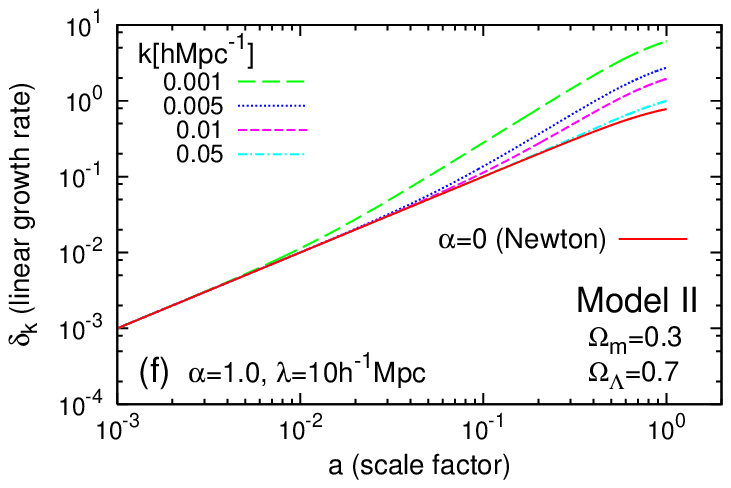}} 
    \end{tabular}
\caption{\label{fig:fig2} Linear growth rate in Model II;
(a) $\alpha$-dependence for $\lambda=10h^{-1}$Mpc and
 $k=0.01h$Mpc$^{-1}$ ($\Omega_{\rm m}=1$ and $\Omega_\Lambda=0$),
(b) $\lambda$-dependence for $\alpha=1$ and
 $k=0.01h$Mpc$^{-1}$ ($\Omega_{\rm m}=1$ and $\Omega_\Lambda=0$),
(c) $k$-dependence for $\alpha=1$ and
 $\lambda=10h^{-1}$Mpc  ($\Omega_{\rm m}=1$ and $\Omega_\Lambda=0$),
(d) $\alpha$-dependence for $\lambda=10h^{-1}$Mpc and
 $k=0.01h$Mpc$^{-1}$ ($\Omega_{\rm m}=0.3$ and $\Omega_\Lambda=0.7$),
(e) $\lambda$-dependence for $\alpha=1$ and
 $k=0.01h$Mpc$^{-1}$ ($\Omega_{\rm m}=0.3$ and $\Omega_\Lambda=0.7$),
(f) $k$-dependence for $\alpha=1$ and
 $\lambda=10h^{-1}$Mpc  ($\Omega_{\rm m}=0.3$ and $\Omega_\Lambda=0.7$).
In all the panels, the results for $\alpha=0$ (Newton's law of gravity)
 are plotted in solid curves for reference.}
\end{figure*}

\section{Linear perturbation theory}\label{Linear}

In this section, we focus on linear evolution of density fluctuation and
solve Eq. (\ref{eq:basiceq}) for the two models. Let us begin with Model
I.  In this case, the Fourier transform of Eq. (\ref{eq:ModelI}) in the
comoving coordinate yields
\begin{equation}
\label{eq:pkModelI}
\ldka \Delta_\bx \Phi ( \bx ) \rdka_{\bk}
	= 4 \pi G a^2 \rb \lka 1 + \alpha_0 \rka \delta_{\bk},
\end{equation}
where $a$ is the scale factor (normalized to unity at $z=0$) and
$\delta_{\bk}$ denotes the Fourier transform of $\delta$.
Note that $\bx$ is the comoving coordinate and we define $\bk$ as the
comoving wave-number.

Substituting Eq. (\ref{eq:pkModelI}) into Eq. (\ref{eq:basiceq}), we
obtain
\begin{equation}
\label{eq:eqModelI}
\ddot{\delta}_\bk + 2 H \dot{\delta}_\bk
	- \frac{3}{2}  ( 1 + \alpha_0 ) 
	\lka 1 - \frac{H_0^2 \Omega_{\Lambda}}{H^2} \rka H^2 \dbk = 0
\end{equation}
in spatially flat models, where $H_0$ is the Hubble constant at the present.

In particular, we find that Eq. (\ref{eq:eqModelI}) has an analytic
solution in the Einstein-de Sitter model:
\begin{equation}
\label{eq:soModelI}
\dbk \propto a^{- \frac{1}{4} \pm \frac{\sqrt{1 + 24 ( 1 +
 \alpha_0)}}{4}}. 
\end{equation}
 Figure \ref{fig:fig1} plots the linear growth rate in Model I as a
function of $a$ which is independent of $k$.  The amplitude in this
figure is normalized $10^{-3}$ at $a=10^{-3}$.

 Next consider Model II.  In Fourier space, the potential is written
as
\begin{equation}
\label{eq:pkModelII}
\ldka \Delta_{\bx} \Phi ( \bx ) \rdka_{\bk}
	= 4 \pi G_{\rm N} a^2 \rb \ldka 1 + \alpha 
	 \frac{\lka \frac{a}{k \lambda} \rka^2}{1 + \lka \frac{a}{k
	 \lambda} \rka^2} \rdka \dbk,
\end{equation}
The linear perturbation equation for density fluctuation can be written
as
\begin{eqnarray}
\label{eq:eqModelII}
 \ddot{\delta}_\bk &+& 2 H \dot{\delta}_\bk \cr 
&-& \frac{3}{2} \ldka 1 + \alpha 
	\frac{\lka \frac{a}{k \lambda} \rka^2}{1 + \lka \frac{a}{k
	\lambda} \rka^2} \rdka 
	\lka 1 - \frac{H_0^2 \Omega_{\Lambda}}{H^2} \rka H^2 \dbk = 0.~~
\end{eqnarray}
We also find an exact analytic solution for Eq. (\ref{eq:eqModelII}) in
the Einstein-de Sitter model:
\begin{widetext}
\begin{align}
\label{eq:soModelII}
	\dbk (a; \alpha, \lambda) &= 
	C_1 \frac{a}{k \lambda} \;_2 F_1 \lka 
	\frac{5}{8} - \frac{1}{8} \sqrt{25+24 \alpha},
	\frac{5}{8} + \frac{1}{8} \sqrt{25+24 \alpha},
	\frac{9}{4}, -\lka \frac{a}{k \lambda} \rka^2 \rka \cr
	\quad & +
	C_2 \lka \frac{a}{k \lambda} \rka^{-3/2} \;_2 F_1 \lka 
	-\frac{5}{8} - \frac{1}{8} \sqrt{25+24 \alpha},
	-\frac{5}{8} + \frac{1}{8} \sqrt{25+24 \alpha},
	-\frac{1}{4}, -\lka \frac{a}{k \lambda} \rka^2 \rka,
\end{align}
\end{widetext}
where $\;_2 F_1$ is the hypergeometric function, and the two constants
$C_1$, $C_2$ are determined by initial conditions.  The first and 
second terms of the right-hand side in Eq.(\ref{eq:soModelII})
correspond to the growing and decaying modes, respectively.

Note that Sealfon et al. \cite{Sealfon} introduced $s \equiv a/k
\lambda$ and attempted to solve Eq. (\ref{eq:eqModelII}) for
$\hat{d}(\bk, t)$:
\begin{equation}
\label{eq:eqSealfon}
	\delta(\bk,t) = \delta_A (\bk,t) \ldka 1 + \alpha \hat{d}(\bk,t)
	 \rdka,
\end{equation}
where $\delta_A(\bk,t)$ denotes the growing mode of density fluctuations
for the $\alpha = 0$ case.  In doing so they neglect higher-order terms of
$\alpha$ and assume that $\hat{d}(\bk,t)$ can be written entirely as a
function of $s$.  Their scaling assumption is correct only in the
Einstein-de Sitter model, but does not hold in general since the Hubble
parameter in Eq. (\ref{eq:eqModelII}) depends on $a$ and thus $\hat{d}$
cannot be expressed only in terms of $s$.  Therefore we chose to
 solve Eq. (\ref{eq:eqModelII}) directly, without neglecting higher-order
terms of $\alpha$.

 Figure \ref{fig:fig2} shows the linear growth rate in Model II as a
function of $a$, which is similar to Fig. \ref{fig:fig1}; the left and
right panels correspond to $\Omega_{\rm m}=1$ and $\Omega_\Lambda=0$ and
to $\Omega_{\rm m}=0.3$ and $\Omega_\Lambda=0.7$, respectively.  In this
case, the result depends on $k$ and $\lambda$ as well, and we show the
$\alpha$-, $\lambda$-, and $k$-dependence in the top, middle, and bottom
panels.

 We emphasize here that $\lambda$ is defined in proper coordinates.
Hence the gravity on the comoving scale corresponding to $k$ starts to
deviate from the conventional Newton potential after $a > k \lambda$.
This is why all of the curves in Fig.\ref{fig:fig2} are degenerate for
$a \le 0.01$, and also explains the fact that the linear growth rate for
smaller $k$ deviates earlier from the $\alpha = 0$ case in the bottom
panels.

\section{Correction for gravitational non-linear growth}
\label{Non-linear}

Our next task is to apply a correction for gravitational non-linearity
to the linear predictions described in the previous section.  The linear
power spectrum in the CDM models extrapolated to $z=0$ ($a=1$ in our
definition) is computed according to
\begin{equation}
\label{eq:PL}
	P_{L}(k; \alpha, \lambda) = A T^2(k) k^{n}
 [\delta_{\bk} (a=1; \alpha,\lambda)]^2
\end{equation}
where $T(k)$ is the matter transfer function and $k \equiv |\bk|$.  In
the above we assume that the primordial spectrum obeys a power-law and
fix the value of $n$ to be unity.  In practice we use the fitting
formula \cite{Eisenstein:1997ik} for $T(k)$ which takes into account the
baryon effect (we adopt $\Omega_{\rm b} = 0.04$).  It should be
emphasized here that we choose the amplitude $A$ so that the rms value
of the top-hat mass fluctuation at $8h^{-1}$Mpc equals those listed in
Table I {\it when $\alpha=0$ and $n=1$}. Strictly speaking, however, our
modified ($\alpha \not= 0$ and $\lambda \not=0$) gravity models distort
the linear spectrum due to the factor $[\delta_{\bk} (a=1;
\alpha,\lambda)]^2$ in Eq. (\ref{eq:PL}). Hence the actual value of
$\sigma_8$ in each model is slightly different from that in the
$\alpha=0$ case. This will be discussed in detail in Sec. V (see
Fig.\ref{fig:fig6} below).

Following the Peacock-Dodds prescription, we compute the nonlinear
counterpart of Eq. (\ref{eq:PL}).  For that purpose, it is convenient to
define
\begin{equation}
\label{eq:daftest}
\Delta_{\rm L}^2(k; \alpha, \lambda) 
\equiv 4 \pi k^3 P_{\rm L}(k; \alpha, \lambda). 
\end{equation}
Then using their nonlinear mapping \cite{Peacock:1996ci}, we transform
$(k_{\rm L}, \Delta^2_{\rm L}(k_{\rm L}; \alpha, \lambda))$ to $(k_{\rm
NL}, \Delta^2_{\rm NL}(k_{\rm NL}; \alpha, \lambda))$, where subscripts
$L$ and $NL$ refer to variables in linear theory and to those corrected
for gravitational non-linear growth, respectively.  In what follows, we
designate $\Delta^2_{\rm L}$ and $\Delta^2_{\rm NL}$ as the linear and
non-linear power spectra.

\begin{figure}
\begin{tabular}{c}
      \resizebox{80mm}{!}{\includegraphics{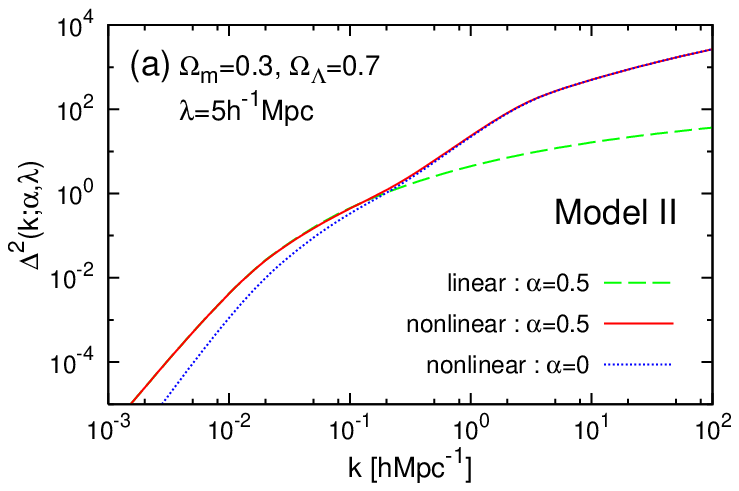}} \\
      \resizebox{80mm}{!}{\includegraphics{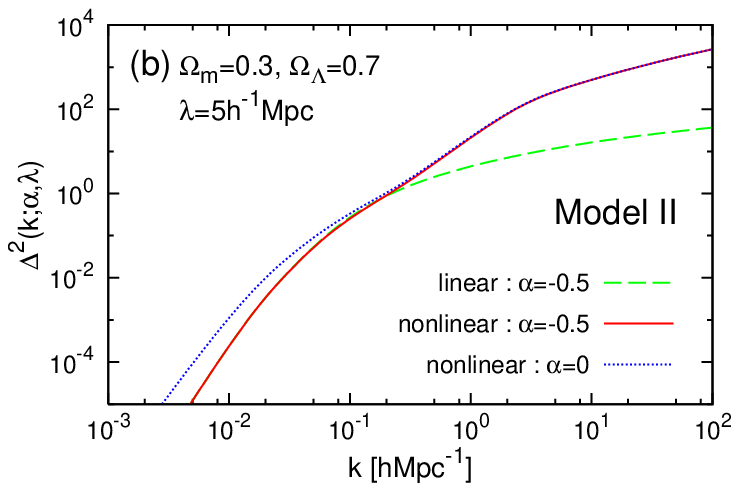}}
\end{tabular}
\caption{\label{fig:fig3} Nonlinear power spectra in Model II; (a)
$\alpha=0.5$, (b) $\alpha=-0.5$. We adopt $\Lambda$-CDM and
$\lambda=5h^{-1}$Mpc for definiteness.  Dashed and solid curves indicate
the linear and nonlinear predictions as discussed in the text.  For
reference, nonlinear Newtonian predictions ($\alpha=0$) are also shown
in dotted curves.}
\end{figure}

Consider Model II first.  Figure \ref{fig:fig3} plots the linear and
non-linear power spectra for $\lambda = 5 h^{-1} {\rm Mpc}$ in the
$\Lambda$-CDM model with $\alpha = 0.5$ and $-0.5$.  Dashed and solid
curves indicate $\Delta_{\rm L}^2(k; \alpha, \lambda)$ and $\Delta_{\rm
NL}^2(k; \alpha, \lambda)$, and $\Delta_{\rm NL}^2(k; \alpha=0,
\lambda)$ is plotted in dotted curves for reference.  Since all our
models have $\sigma_8 \sim 1$ (Table I), the non-linearity shows up
around wave-numbers $k \sim 2 \pi/(8 h^{-1} {\rm Mpc})$.  As expected,
the effect of modified gravity becomes significant for $k \ll 2
\pi/ \lambda$.

\begin{figure}
\begin{tabular}{c}
      \resizebox{80mm}{!}{\includegraphics{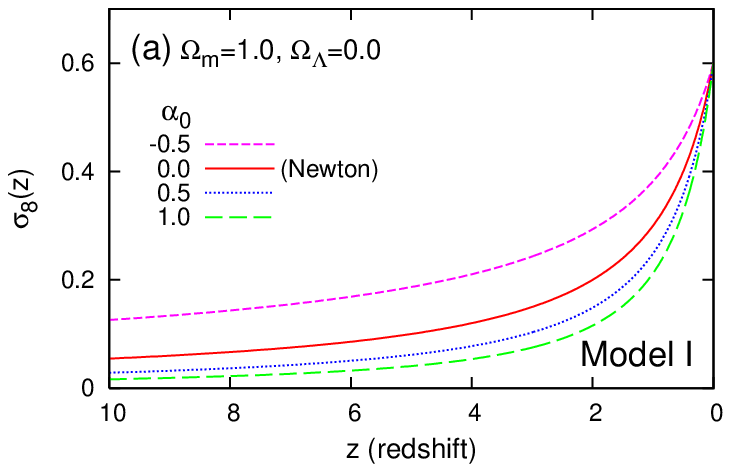}} \\
      \resizebox{80mm}{!}{\includegraphics{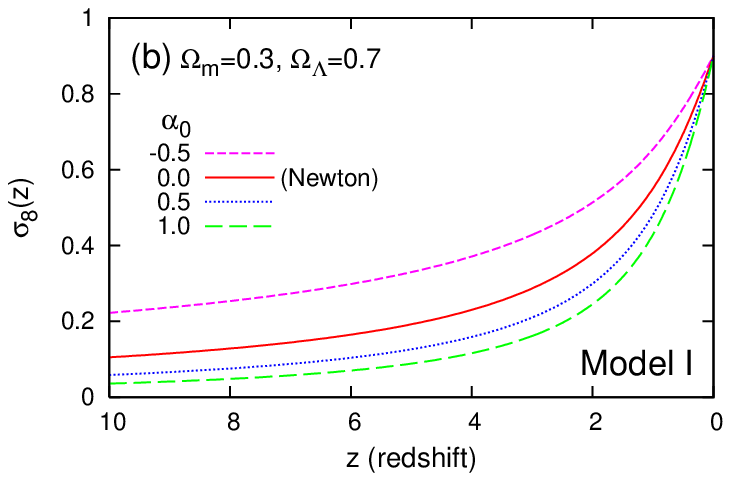}}
\end{tabular}
\caption{\label{fig:fig4} Linearly extrapolated values of
$\sigma_8(z)$ for different $\alpha_0$ in Model I; (a)
$\Omega_{\rm m}=1.0$ and $\Omega_\Lambda=0.0$, 
(b) $\Omega_{\rm m}=0.3$ and $\Omega_\Lambda=0.7$.
}
\end{figure}

In contrast, Model I does not change the nonlinear power spectrum at
$z=0$ relative to the $\alpha_0=0$ case {\it if} one adopts the
identical value of $\sigma_8$ at $z=0$.  Since the linear growth rate
for Model I (Eq. (\ref{eq:soModelI})) is independent of $k$, the above
statement is obvious for linear power spectra. The possible difference
would come from the nonlinear evolution. If one evolves $\alpha_0=0$ and
$\alpha_0 \not= 0$ model spectra from an earlier epoch with the
identical amplitude (i.e., $\sigma_8$ at that epoch), they lead to
different values of $\sigma_8$ at $z=0$ because the $\alpha_0$-dependent
overall growth rate, and thus to different shape of the power spectra
due to the nonlinear effect. Instead, if one chooses the identical value
of $\sigma_8$ at $z=0$, the power spectra of $\alpha_0=0$ and $\alpha_0
\not= 0$ models are also identical even taking into account the
nonlinear effect.  To disentangle Model I from the $\alpha_0 = 0$ model,
therefore, one has to compare $\Delta_{\rm NL}^2$ at different
redshifts. Figure \ref{fig:fig4} displays the value of $\sigma_8$ as a
function of $z$. Since this value is computed by extrapolating linear
theory, it can be directly obtained from the linear growth rate plotted
in Fig. \ref{fig:fig1}. Thus in Model I the structure evolves at a
different rate even if the underlying cosmological parameters
($\Omega_{\rm m}$ and $\Omega_\Lambda$) are identical. For example, we
would need determinations of $\sigma_8$ at multiple redshifts to resolve
the degeneracy.

\label{sec:Constraints}
\section{Comparison with the power spectrum of SDSS galaxies}

We are now in a position to explore observational constraints on
deviations from Newton's law of gravity.  In order to compare the model
predictions described in the previous section against observations, we
use the power spectrum of $L_*$ galaxies derived from SDSS by Tegmark et
al. \cite{Tegmark} (their Table 3 and
http://www.hep.upenn.edu/{\textasciitilde}max/sdss.html); they compute
the power spectrum in the range $0.02 h {\rm Mpc}^{-1} < k < 0.3 h {\rm
Mpc}^{-1}$ from a sample of 205,443 galaxies with mean redshift $z \sim
0.1$.  Their measurements are in good agreement with our $\Lambda$-CDM
model listed in Table I as long as $\alpha = 0$.

\begin{figure}
\begin{tabular}{c}
	\resizebox{80mm}{!}{\includegraphics{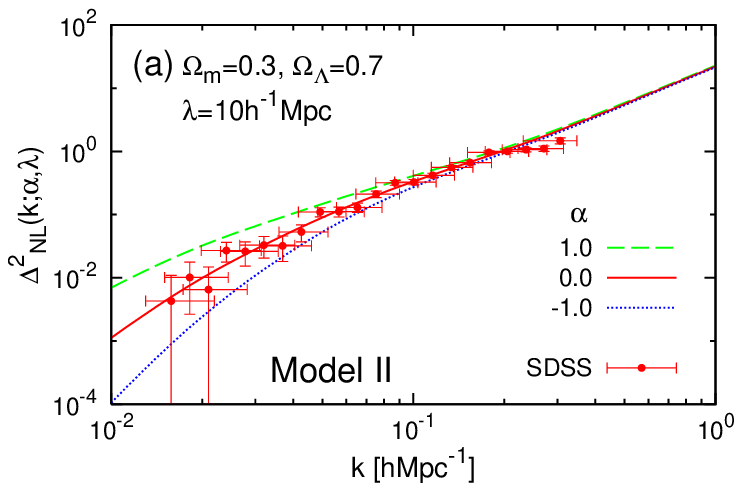}} \\
      \resizebox{80mm}{!}{\includegraphics{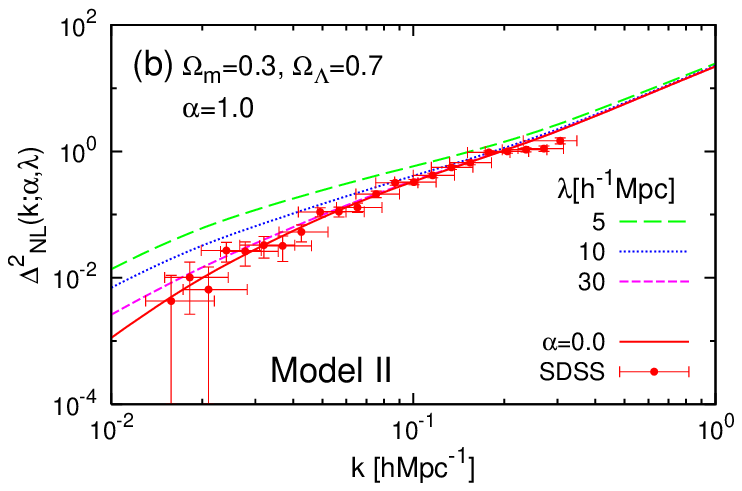}}
\end{tabular}
\caption{\label{fig:fig5} Comparison of power spectra 
between Model II predictions 
and the SDSS galaxies (in the $\Lambda$-CDM model);
(a)  $\alpha = 1$ (dotted) and $-1$ (long-dashed) for 
$\lambda=10 h^{-1}$Mpc, and
(b) $\lambda=5 h^{-1}$Mpc (long-dashed), $10 h^{-1}$Mpc (dotted), 
and  $30 h^{-1}$Mpc (short-dashed)
for $\alpha=1$.
For reference, the Newtonian case ($\alpha=0$) is plotted in solid curves.
The data points with errors bars represent the results of the SDSS
 galaxies.
}
\end{figure}

Figure \ref{fig:fig5} compares non-linear power spectra for Model II
against those for the SDSS galaxies in the $\Lambda$-CDM model; Figure
\ref{fig:fig5}(a) shows the $\alpha$-dependence for the $\lambda = 10
h^{-1} {\rm Mpc}$ case, while Fig. \ref{fig:fig5}(b) illustrates that of
varying $\lambda$ for $\alpha = 1.0$.

\begin{figure}
\begin{tabular}{c}
      \resizebox{80mm}{!}
{\includegraphics{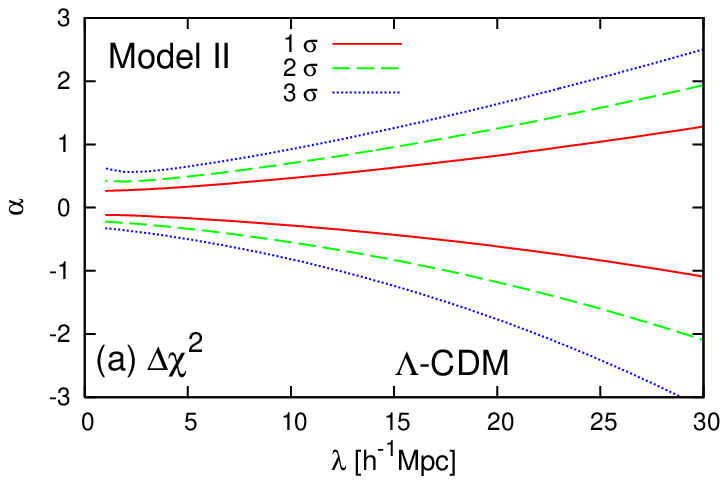}} \\
      \resizebox{80mm}{!}
{\includegraphics{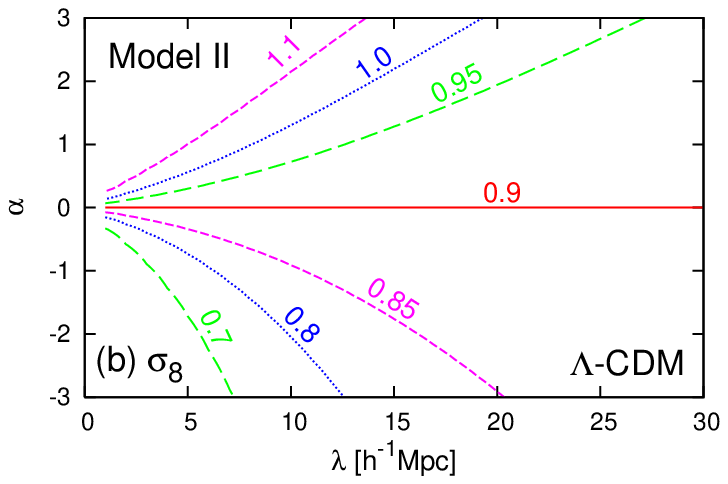}}\\
      \resizebox{80mm}{!}
{\includegraphics{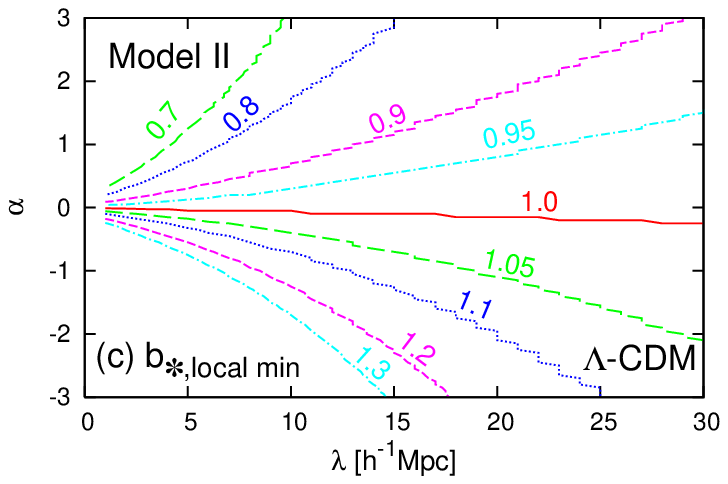}}
\end{tabular}
\caption{\label{fig:fig6}
Constraints in the $\alpha$-$\lambda$ plane for Model II in $\Lambda$-CDM
cosmology;
(a) contours of $\Delta \chi^2$,
(b) contours of $\sigma_8$, 
(c) contours of $b_{*, \rm local~min}$.
}
\end{figure}

In order to proceed further, we constrain $\alpha$ and
$\lambda$ by applying $\Delta \chi^2$ statistics \cite{press}.  Model
II has two independent parameters ($\alpha$ and $\lambda$), and the
overall normalization between the predictions and the data is adjusted
by the biasing parameter $b_*$ for $L_*$ galaxies which we adopt as a
third fitting parameter.  We then compute the relative confidence levels
with respect to the best-fit values assuming that
\begin{eqnarray}
\label{eq:chi}
\Delta \chi^2 (\alpha,\lambda) &\equiv&
	\chi^2(\alpha,\lambda,b_{*, \text{local~min}}) \cr
 && - \chi^2(\alpha_{\rm min},\lambda_{\rm min},b_{*, \text{min}}) 
\end{eqnarray}
follows the $\chi^2$distribution for 2 degrees of freedom.  In
Eq.(\ref{eq:chi}), $\alpha_{\rm min}$, $\lambda_{\rm min}$ and $b_{*,
\text{min}}$ denote their best-fit values which globally minimize the
value of $\chi^2$, while $b_{*,\text{local~min}}$ is the value that
minimizes the $\chi^2$ for a given set of values for $\alpha$ and
$\lambda$.  The results are summarized in Fig.\ref{fig:fig6}; Figure
\ref{fig:fig6}(a) plots the contours of $\Delta \chi^2
(\alpha,\lambda)$; solid, long-dashed and dotted lines indicate the
limits at $1\sigma$, $2\sigma$, and $3\sigma$ confidence levels.  The
amplitude of the power spectrum predicted for Model II is chosen so as
to lead to $\sigma_8=0.9$ at $z=0$ {\it when} $\alpha=0$.  Thus using
the linear growth rate plotted in Fig. \ref{fig:fig2}, one can compute
the contours of $\sigma_8$ for Model II in the $\alpha$-$\lambda$ plane
which are shown in Fig. \ref{fig:fig6}(b).  Figure \ref{fig:fig6}(c)
indicates the corresponding $b_{*, \text{local~min}}(\alpha,\lambda)$
for the SDSS galaxies with respect to each model.  If the linear growth
rate were exactly scale-independent and the fit were performed only over
the fully linear regime, $\sigma_8 b_{*,
\text{local~min}}(\alpha,\lambda)$ plotted in Figs.\ref{fig:fig6}(b) and
(c) would be constant.  In reality, however, the linear growth rate is
slightly scale-dependent and the range which we fix, $0.02 h {\rm
Mpc}^{-1} < k < 0.3 h {\rm Mpc}^{-1}$, spans a (mildly) nonlinear regime
as well. Thus $\sigma_8 b_{*, \text{local min}}(\alpha,\lambda)$ is not
exactly the same over the entire $\alpha$-$\lambda$ plane.

\label{sec:Discussion} 
\section{Conclusion and Discussion}

 We have used a simple empirical model to parameterize the departure
from the Newtonian law of gravity on cosmological scales.  We have
computed the nonlinear power spectra of density fluctuations predicted
in the model assuming a spatially-flat cosmological model and derived
the constraints on the amplitude $\alpha$ and the length scale $\lambda$
which characterize the additional Yukawa-type term.  Our results are an
improvement over the previous attempt by Sealfon et al. as discussed in
Secs. I and III.  For instance, we obtain the 99.7\% confidence limits
of $-0.5<\alpha <0.6$ and $-0.8<\alpha<0.9$ for $\lambda=5h^{-1}$Mpc and
$10h^{-1}$Mpc, respectively.

Our Model II, Eq. (\ref{eq:ModelII}), is intended to represent a
reasonable approximation to a wide class of possible models which
recover Newton's gravity on small scales but depart from it on large
scales.  In contrast, Frieman and Gradwohl \cite{Frieman91} and Gradwohl
and Frieman \cite{Gradwohl92} considered a different non-Newtonian
gravity model which deviates from Newton's law on smaller scales, but
approaches it asymptotically on larger scales. While their specific
model of gravity law is different from that studied here, they
extensively discussed the effects on linear perturbation theory and
explored the resulting implications for large-scale structure.  More
recently Nusser, Gubser and Peebles \cite{Nusser04} considered
cosmological constraints on a similar model of gravity which deviates on
small scales. In order to avoid conflict with laboratory tests of
Newton's law, they have to assume that non-Newtonian gravity applies
only to dark matter; their model is based on the proposal of
Ref. \cite{Farrar-Peebles} in which the dark sector consists of two
mutually coupled fields (dark matter and dark energy), and the length
scale $\lambda$ is constant in {\it comoving} coordinate.  This is why
their model substantially changes the nonlinear clustering behavior
while our model mostly modifies the weakly nonlinear regime. 

In either case, more specific models of non-Newtonian gravity are needed
in order to tighten the constraints. Possible proposals include the DGP
model in the context of the braneworld scenario
\cite{Dvali:2000hr,Deffayet:2000uy}, and a relativistic version of the
modified Newtonian dynamics \cite{Bekenstein04}. These examples are
attractive in the sense that they offer specific modified gravity laws
on local scales and modified Friedmann equations simultaneously in a
consistent fashion.  Therefore we could derive more stringent
constraints.

Using the SDSS galaxy power spectrum is appropriate for constraining the
deviation around $10 \sim 100 h^{-1}$Mpc scales. If a deviation from
Newton's law occurs at scales below $1h^{-1}$Mpc, strong gravitational
nonlinearity will be important, and nonlinear spherical collapse
analysis and/or direct N-body simulations would be needed to explore the
dynamical consequences for dark halos at galaxy and cluster scales.  If
the deviation scale is much larger than $100h^{-1}$Mpc, on the other
hand, it is unlikely to leave any detectable signature on cosmic
structures.

\section*{Acknowledgements}

We would like to thank Akio Hosoya, Shinji Mukohyama, Masaru Siino, and
Atsushi Taruya for useful discussions and comments.  We are also
grateful to Josh Frieman for calling our attention to very relevant
previous literature, and to Ed Turner for a careful reading of the
manuscript and for numerous suggestions.  This work was partially
supported by Grants-in-Aid for Scientific Research from the Japan
Society for Promotion of Science (Nos.13135208, 14740155 and 14102004,
16340053).

\appendix

\end{document}